\documentclass[conference,letterpaper]{IEEEtran}

\usepackage[utf8]{inputenc}
\usepackage[T1]{fontenc}
\usepackage{url}
\usepackage{ifthen}
\usepackage{cite}
\usepackage[cmex10]{amsmath} 

\RequirePackage{bm}

\RequirePackage{amssymb}
\RequirePackage{amsfonts}
\RequirePackage{amsthm}
\RequirePackage{mathtools}
\RequirePackage{dsfont}
\RequirePackage{cases}
\RequirePackage{float}

\RequirePackage[boxed,ruled,vlined]{algorithm2e}
\RequirePackage{algorithmic}

\RequirePackage{enumitem}

\RequirePackage{subfigure}

\RequirePackage{epsfig}
\RequirePackage{epstopdf}

\makeatletter\def\CT{\def\@captype{figure}}\makeatother

\RequirePackage{booktabs}
\RequirePackage{rotating}
\RequirePackage{lscape}
\RequirePackage{pdflscape}
\RequirePackage{longtable}

\makeatletter
\def\th@plain{%
	\thm@notefont{}
	\itshape 
}
\def\th@definition{%
	\thm@notefont{}
	\normalfont 
}
\makeatother

\theoremstyle{plain}

\newtheorem{theorem}{Theorem}

\newtheorem{lemma}{Lemma}

\theoremstyle{definition}

\newtheorem{definition}{Definition}

\newtheorem{remark}{Remark}

\newtheorem{fact}{Fact}

\usepackage{mysty}

\begin{document}
\title{Time-Data Tradeoffs in Structured Signals Recovery via the Proximal-Gradient Homotopy Method}


\author{%
	\IEEEauthorblockN{Xiao~Lv and Wei~Cui}
	\IEEEauthorblockA{School of Information and Electronics\\
		Beijing Institute of Technology\\
		Beijing 100081, China\\
		Email: \ba{xiaolv, cuiwei}@bit.edu.cn}
	\and
	\IEEEauthorblockN{Yulong~Liu}
	\IEEEauthorblockA{School of Physics\\
		Beijing Institute of Technology\\
		Beijing 100081, China\\
		Email: yulongliu@bit.edu.cn}
}


\maketitle

\begin{abstract}
In this paper, we characterize data-time tradeoffs of the proximal-gradient homotopy method used for solving linear inverse problems under sub-Gaussian measurements. Our results are sharp up to an absolute constant factor. We demonstrate that, in the absence of the strong convexity assumption, the proximal-gradient homotopy update can achieve a linear rate of convergence when the number of measurements is sufficiently large. Numerical simulations are provided to verify our theoretical results.
\end{abstract}


\section{Introduction}
Throughout science and engineering, one is often faced with the problem of recovering a structured signal from a relatively small number of linear measurements
\begin{equation}\label{Problem}
	\bmy = \bmA \xreal + \bmomega,
\end{equation}
where $\bmA \in \RR^{m \times n}$ is the sensing matrix with $m \leq n$, $\xreal \in \RR^{n}$ denotes the desired structured signal, and $\bmomega\in \RR^{m}$ stands for the random noise. The objective is to recover $\xreal$ from given knowledge of $\bmy$ and $\bmA$.

Since the problem is generally ill-posed, tractable recovery is achievable when signal is well structured. Typical structured signals include sparse vectors and low-rank matrices. Let $\Reg{\cdot}$ be a suitable norm which promotes the structure for signal. There are three popular convex recovery procedures to reconstruct signal when different kinds of prior information are available. Specifically, when $\Reg{\bmx^{\star}}=R$ or the noise level $\delta$ (in terms of the $\ell_2$-norm) is available, it is natural to consider the following constrained convex recovery procedure
\begin{equation}\label{Problem_constrained}
	\begin{split}
		&\umin{\bmx} \frac{1}{2} \ltwonorm{\bmA \bmx - \bmy}^2 \quad \st \quad \Reg{\bmx} \leq R,
	\end{split}
\end{equation}
or
\begin{equation}\label{Problem_constrained2}
	\begin{split}
		&\umin{\bmx} \Reg{\bmx}  \quad \st \quad \ltwonorm{\bmA \bmx - \bmy} \leq \delta.
	\end{split}
\end{equation}
When there is no prior knowledge available, it is practical to use the penalized recovery procedure
\begin{equation}\label{Problem_reg}
	\umin{\bmx} \quad \lambda \Reg{\bmx} + \frac{1}{2} \ltwonorm{\bmA \bmx - \bmy}^2,
\end{equation}
where $\lambda>0$ is a tradeoff parameter.

To obtain the original signal in practice, we need to solve the above recovery procedures by some specific algorithms. Typical examples include first-order optimization techniques (e.g., ISTA \cite{Blumensath2008ISTA}, FISTA \cite{Beck2009AFI}, ADMM \cite{Boyd2010DistributedO}, and NESTA \cite{Becker2011NESTA}) and second-order optimization techniques (e.g., the interior-point method \cite{Kim2007AnItnteriorP} and Newton-like methods \cite{Meng2020NewtonStepB,Zhou2021GlobalandQ}).


For the problem of structural signals recovery, a fundamental problem is to characterize the relationships among data complexity (or the number of measurements), structural complexity (or the structure of the desired signal), and time complexity (or the convergence rate of certain algorithm). During the past decade, there are a large number of works studying data-structure tradeoffs, see e.g., \cite{Chandrasekaran2012TheCG},\cite{Amelunxen2013Living},\cite{Tropp2015ConvexRecovery},\cite{Thrampoulidis2015RecoveringStructured},\cite{Vershynin2015EstimationInHigh},\cite{Vaiter2015LowComplexity}, and references therein. These results reveal that for a given recovery procedure, how many measurements (data) are required to guarantee successful recovery of a structured signal. The minimal number of measurements to guarantee recovery is usually determined by the structure of the original signal, so this relationship is called data-structure tradeoffs.

Much less work is devoted to studying tradeoffs between time complexity and data complexity, see e.g.,\cite{chandrasekaran2013computational},\cite{bruer2015designing}\cite{oymak2018sharp},\cite{Oymak2017FastR}. In \cite{oymak2018sharp}, Oymak, Recht, and Soltanolkotabi have established sharp time-data tradeoffs in solving the constrained optimization problem \eqref{Problem_constrained} via the projected gradient descent (PGD) algorithm. In real-word applications, however, it is more practical to use the penalized recovery procedure \eqref{Problem_reg} since it requires no prior information. In \cite{Oymak2017FastR}, Oymak and Soltanolkotabi have tried to establish time-data tradeoffs in solving the penalized optimization problem \eqref{Problem_reg} via the proximal-gradient homotopy scheme. Their analysis depends on a very impractical resampling assumption, namely, independent copies of both observations $\bmy$ and the sensing matrix $\bmA$ are used in each iteration for their convergence analysis. This paper tries to remove the assumption and to solve an important problem suggested in \cite{Oymak2017FastR, oymak2018sharp}. Specifically, we establish time-data tradeoffs (without resampling assumption) in solving the penalized optimization problem \eqref{Problem_reg} via the proximal-gradient homotopy method for three kinds of typical structured signals, namely, sparse vectors, group sparse vectors and low-rank matrices.

\section{Preliminaries}
In this section, we introduce some notations that underlie our analysis. Throughout this paper, $\mathbb{S}^{n-1}$ denotes the unit sphere in $\RR^n$ under the $\ell_2$-norm.

\begin{definition}[Sub-Gaussian random variables and vectors]
	A random variable $X$ is called a sub-Gaussian random variable if its Orlicz norm
	\begin{equation}
		\subgnorm{X} = \inf \ba{t > 0 \colon \EE \psi_2(\abs{X} / t) \leq 2} \label{SubGnorm}
	\end{equation}
is finite for $\psi_2(x) = \exp(x^2) - 1$. The sub-Gaussian norm of $X$ is defined to be the smallest $t$ in (\ref{SubGnorm}), donated by $\subgnorm{X}$.

A random vector $\bmx$ in $\RR^n$ is sub-Gaussian if all of its one-dimensional marginals are sub-Gaussian random variables and its $\psi_2$-norm is defined as
	\begin{equation*}
			\subgnorm{\bmx} = \usup{\bmu \in \mathbb{S}^{n - 1}} \subgnorm{\iprod{\bmx}{\bmu}}.
	\end{equation*}
\end{definition}
\begin{definition}[Gaussian complexity and Gaussian width]
	For any $\calC \subseteq \RR^n$, the Gaussian complexity is a simple way to quantify its size
		\begin{equation*}
			\gcomp{\calC} \coloneqq \EE \usup{\bmx \in \calC} \abs{\iprod{\bmg}{\bmx}},
		\end{equation*}
	where $\bmg \sim \calN(\bmzero, \bmI_n)$. A closely related geometric quantity is the Gaussian width
		\begin{equation*}
			\gwidth{\calC} \coloneqq \EE \usup{\bmx \in \calC} \iprod{\bmg}{\bmx},
		\end{equation*}
	where $\bmg \sim \calN(\bmzero, \bmI_n)$.
\end{definition}

We also define the following two restricted singular values which play a key role in our analysis. Let $\calP \subseteq \mathbb{S}^{n - 1}$ and $\calQ \subseteq \mathbb{S}^{n - 1}$,
\begin{align}
	\rho(\calP,\calQ) &= \usup{\bmv \in \calP, \bmu \in \calQ} \iprod{\bmv}{(\bmI - \mu \bmA^{T} \bmA)\bmu} \label{restricted_singular_values1}\\
	\xi(\calP) &= \usup{\bmv \in \calP} \iprod{\bmv}{\bmA^{T} \frac{\bmomega}{\ltwonorm{\bmomega}}} \label{restricted_singular_values2}  .
\end{align}


\section{Time-Data Tradeoffs in Sparse Signals Recovery} \label{Sparse}
When the desired signal is sparse, it is natural to use the $l_1$-norm as the regularizer, i.e., $\Reg{\bmx} = \|\bmx\|_1$. To recover the original signal, we use the proximal-gradient homotopy method to solve the penalized optimization problem \eqref{Problem_reg}. Specifically, starting from an initial point $\bmx_0$ (often set to $\mathbf{0}$), the proximal-gradient homotopy algorithm proceeds as follows \cite{Beck2017FirstOM}:
\begin{equation}\label{Proximal_sparse}
	\begin{split}
		\xtp & = \Prox{\yt}{ {\lambdat}{\mu} \Reg{\cdot}}= \Prox{\yt}{ {\lambdat}{\mu} \|\cdot\|_1} \\
		     & = \uargmin{\bmx}  \left \{ {\lambdat}{\mu} \lonenorm{\bmx} + \frac{1}{2}\ltwonorm{\bmx - \yt}^2 \right\},
	\end{split}
\end{equation}
where $\lambdat$ is the homotopy continuation parameter, $\mu$ denotes the step size, and $\yt = \xt - \mu   \nabla (\|\bmA\bmx_t - \bmy\|_2^2/2) = \xt -  \mu \bmA^T(\bmA \xt -\bmy)$. In particular, if the prior information of $\|\bmx^{\star}\|_1$ is available, then we can define $\calK =\{\bmx ~|~ \|\bmx\|_1 \leq R = \|\bmx^{\star}\|_1\} $ and its corresponding indicator function
$$\iota_{\calK}(\bmx) =   \left\{
\begin{matrix}
0 ~~~\textrm{if} ~ \bmx \in \calK,\\
\infty~~~ \textrm{otherwise}. \\
\end{matrix}
\right.$$
It is not hard to find that letting $\Reg{\cdot} = \iota_{\calK}(\cdot) $ in \eqref{Proximal_sparse} will lead to the standard projected gradient descent (PGD) algorithm:
\begin{equation}\label{PDG}
  \bmx_{t+1} = \mathcal{P_{\calK}} (\xt -  \mu \bmA^T(\bmA \xt -\bmy)),
\end{equation}
where $\calP_{\calK}(\bmx)$ denotes the Euclidean projection of $\bmx$ on to the set $\calK$.

Sharp time-data tradeoffs in solving the constrained optimization problem \eqref{Problem_constrained} via PGD \eqref{PDG} are well established in \cite{oymak2018sharp}. A key ingredient in establishing the linear convergence rate for PGD is that each iteration $\bmx_t$ is restricted in a structural constraint set, i.e., $ \bmx_t  \in \calK$. In the proximal-gradient homotopy scheme, however, there is no such explicit structural constraint set in general. In order to overcome this difficulty, we carefully choose the continuation parameter $\lambda_t$ such that each iteration of \eqref{Proximal_sparse} belongs to a certain structural constraint set.

\begin{lemma}[Implicit constraint induced by $\lambda_t$] \label{Constraint_sparse}
    Consider the observation model \eqref{Problem} with $\lzeronorm{\xreal} \leq  s$  and $\ltwonorm{\bmomega} \leq \delta$. Suppose $\Card(\supp(\xt) \backslash \supp(\xreal)) < s$ and $\ltwonorm{\bmx_t - \bmx^{\star}} \leq \Delta_t$.
    If we choose $\lambda_{t}$ as
	\begin{equation}
		\lambdat = \frac{ \xi_{s} \delta + \rho_{s,2s} \Delta_t/\mu}{\sqrt{s}}, \label{Lambda_sparse}
	\end{equation}
	then the update $\xtp$ generated by \eqref{Proximal_sparse} satisfies
	\begin{equation*}
		\Card(\supp(\xtp) \backslash \supp(\xreal)) < s,
	\end{equation*}
	where $\calS_s = \ba{\bmx \mid \lzeronorm{\bmx} \leq s}$, $\xi_{s} = \xi(\calS_s \cap \mathbb{S}^{n - 1})$ and $\rho_{s,2s} = \rho(\calS_s \cap \mathbb{S}^{n - 1},\calS_{2s} \cap \mathbb{S}^{n - 1})$.
\end{lemma}

Armed with Lemma \ref{Constraint_sparse}, we then establish time-data tradeoffs in sparse signals recovery via the proximal-gradient homotopy scheme.
\begin{theorem}[Time-data tradeoffs in sparse signals recovery]\label{Convergence_sparse}
Consider the observation model \eqref{Problem} with $\lzeronorm{\xreal} \leq  s$  and $\ltwonorm{\bmomega} \leq \delta$. Suppose the rows of the sensing matrix $\ba{\bmA_i}$ are independent centered isotropic sub-Gaussian vectors with $K = \umax{i} \subgnorm{\bmA_i}$. To estimate $\bmx^{\star}$, we apply the proximal-gradient homotopy update \eqref{Proximal_sparse} with $\mu  = 1/m$ and $\Delta_0 \geq \ltwonorm{\xreal}$. Let the continuation parameter $\lambda_{t}$ be chosen according to (\ref{Lambda_sparse}). If
	\begin{equation}\label{sufficentcondition1}
		\sqrt{m} > CK^2\bPa{\gcomp{\calS_{2s} \cap \mathbb{S}^{n - 1}} + \eta},
	\end{equation}
then each update of \eqref{Proximal_sparse} satisfies
	\begin{equation}\label{Interation_convergence_sparse1}
		\ltwonorm{\bmx_{t+1}-\bmx^{\star}} \leq \rho^{t+1} \Delta_0 + \frac{\xi}{1 - \rho} \frac{\delta}{m}
	\end{equation}
	with probability at least $1 - c\exp\pa{-\eta^2}$. Here
	\begin{align}
		\rho &=  \frac{C'K^2}{\sqrt{m}}\Big(\gcomp{\calS_{2s} \cap \mathbb{S}^{n - 1}} + \eta\Big), \label{ParameterRho1} \\
		\xi  &=  C''K (\gcomp{\calS_{2s} \cap \mathbb{S}^{n - 1}} + \eta),  \label{ParameterXi1}\\
        \Delta_{t + 1} &= \rho \Delta_t + \frac{1}{m} \xi \delta, \label{updateofdelta1}
	\end{align}
and $c, C, C', C''$ are absolute constants.
\end{theorem}

\begin{remark}[Sharpness]
	It follows from \cite[ex. 10.3.8]{vershynin2018HDP} that the Gaussian complexity $\gamma(\calS_s \cap \mathbb{S}^{n - 1})$ can be bounded as $\gcomp{\calS_s \cap \mathbb{S}^{n - 1}} = \gwidth{\calS_s \cap \mathbb{S}^{n - 1}} \lesssim \sqrt{s\log\frac{e n}{s}}$, where the equality holds because of the symmetry of $\calS_s$. This implies that our sufficient condition \eqref{sufficentcondition1} is sharp up to an absolute constant factor.
\end{remark}
\begin{remark}[Linear convergence rate]
	In the absence of noise, i.e., $\delta = 0$, our analysis shows that if the number of measurements is large enough, then the proximal-gradient homotopy update \eqref{Proximal_sparse} has a linear convergence rate. Moreover, \eqref{ParameterRho1} and \eqref{Interation_convergence_sparse1} indicate that more observations will lead to a smaller $\rho$, and hence a faster convergence speed. In addition, in the presence of noise, \eqref{ParameterRho1} and \eqref{Interation_convergence_sparse1} also reveal that more measurements will lead to less estimation error.
\end{remark}

\begin{remark}[Group sparse signals] \label{RemarkGroup}
	It is worth noting that our analysis can be naturally extended to the group sparse signals. Since the results are very similar to the sparse case, we omit them here and only provide the numerical results in Section \ref{Numerical results}.
\end{remark}

\begin{remark}[Related works]
  In \cite{Oymak2017FastR}, Oymak and Soltanolkotabi have tried to establish time-data tradeoffs in solving the penalized optimization problem \eqref{Problem_reg} via the proximal-gradient homotopy scheme. Their analysis depends on a very impractical resampling assumption. More precisely, in each iteration, independent copies of both observations $\bmy$ and the sensing matrix $\bmA$ are used in their analysis. This paper tries to remove this assumption and to solve an important problem suggested in \cite{Oymak2017FastR, oymak2018sharp}. In \cite{agarwal2012Fast}, Agarwal et al. have studied the fast convergence of the proximal-gradient scheme with a constrained proximal mapping, namely, each iteration $\bmx_t$ is restricted in an explicit constraint set. In \cite{Xiao2012ProximalHomotopy} and \cite{Eghbali2017DecomposableNM}, the authors also establish the linear convergence rate for the  proximal-gradient scheme, but their sufficient conditions do not clearly reveal time-data tradeoffs in linear inverse problems.
\end{remark}

Based on the above theoretical results, we summarize the proximal-gradient homotopy update for sparse signals recovery in Algorithm \ref{alg:2}.
\begin{algorithm}[ht]
	\caption{Proximal-gradient homotopy method for sparse signals recovery}
	{\bf Input:} Initial value $\Delta_0$, observation matrix $\bmA$, measurements $\bmy$, maximum sparsity $s$, contraction parameters $\rho$ and $\rho_{s,2s}$, noise parameters $\xi$ and $\xi_{s}$, noise level $\delta$, maximum iteration number $t_{\mathrm{max}}$
	
	\begin{algorithmic}\label{AlgorithmPGH}
		\STATE Initialize $\bmx_0 = \bmzero$
		\FOR{$t=0, 1, \ldots, t_{\mathrm{max}}-1$}
		\STATE $\lambda_{t} = \frac{1}{\sqrt{s}}\xi_{s}\delta + \frac{m}{ \sqrt{s}}\rho_{s,2s} \Delta_t$
		\STATE $\xtp = \Prox{\xt - \frac{1}{m} \bmA^T (\bmA \xt - \bmy)}{\frac{1}{m} \lambda_{t} \lonenorm{\cdot}}$
		\STATE $\Delta_{t+1} = \rho\Delta_t + \frac{1}{m} \xi \delta$
		\ENDFOR
	\end{algorithmic} \label{alg:2}
	{\bf Return} $\bmx_{t_{\mathrm{max}}}$
\end{algorithm}


\section{Time-Data Tradeoffs in Low-Rank Matrices Recovery} \label{Low Rank}
When the desired signal is a low-rank matrix, the observation model becomes
\begin{equation}\label{Problem2}
	\bmy = \calA(\Xreal) + \bmomega,
\end{equation}
where $\calA \in \RR^{m \times d^2}$ is the sensing matrix, $\Xreal \in \RR^{d \times d}$ denotes the targeted low-rank matrix. In this case, we use the nuclear norm as the regularizer, i.e., $\Reg{\bmX} = \|\bmX\|_*$.
We also use the proximal-gradient homotopy method to recover the original signal. Specifically, starting from an initial point $\bmX_0$ (often set to $\mathbf{0}$), the proximal-gradient homotopy algorithm updates as follows:
\begin{equation}\label{Proximal_lowrank}
	\begin{split}
		\Xtp & = \Prox{\Yt}{ {\lambdat}{\mu} \Reg{\cdot}}= \Prox{\Yt}{ {\lambdat}{\mu} \|\cdot\|_*} \\
		     & = \uargmin{\bmX}  \left \{ {\lambdat}{\mu} \|\bmX\|_* + \frac{1}{2}\|\bmX - \Yt\|_F^2 \right\},
	\end{split}
\end{equation}
where $\lambdat$ is the homotopy continuation parameter, $\mu$ denotes the step size, and $\Yt = \Xt - \mu   \nabla (\|\calA(\bmX_t) - \bmY\|_F^2/2) = \Xt -  \mu \calA^T(\calA(\Xt) -\bmY)$. Similar to the sparse case, we choose the continuation parameter $\lambda_t$ such that each iteration of \eqref{Proximal_lowrank} satisfies certain implicit  structural constraint.

\begin{lemma}[Implicit constraint induced by $\lambda_t$ ] \label{Constraint_lowrank}
 Consider the observation model \eqref{Problem2} with $\rank(\Xreal) \leq  r$  and $\ltwonorm{\bmomega} \leq \delta$. Suppose $\rank(\Xt) < 2r$ and $\fnorm{\bmX_t-\bmX^{\star}} \leq \Delta_t$.
    If we choose $\lambda_{t}$ as
	\begin{equation}
		\lambdat = \frac{ \xi_{r} \delta + \rho_{r,3r} \Delta_t/\mu}{\sqrt{r}}, \label{Lambda_lowrank}
	\end{equation}
	then the update $\Xtp$ generated by \eqref{Proximal_lowrank} satisfies
	\begin{equation*}
		\rank(\Xtp) < 2r.
	\end{equation*}
    where $\calS_r = \ba{\bmX \mid \rank{(\bmX)} \leq r}$, $\xi_{r} = \xi(\calS_r \cap \mathbb{S}^{d^2 - 1})$, and $\rho_{r,3r} = \rho(\calS_r \cap \mathbb{S}^{d^2 - 1},\calS_{3r} \cap \mathbb{S}^{d^2 - 1})$.
\end{lemma}

We then establish time-data tradeoffs in low-rank matrices recovery via the proximal-gradient homotopy scheme.

\begin{theorem}[Time-data tradeoffs in low-rank matrices recovery]\label{Convergence_lowrank}
Consider the observation model \eqref{Problem2} with $\rank(\bmX^{\star}) \leq  r$  and $\ltwonorm{\bmomega} \leq \delta$. Suppose the rows of the sensing matrix $\ba{\calA_i}$ are independent centered isotropic sub-Gaussian vectors with $K = \umax{i} \subgnorm{\calA_i}$. To estimate $\bmX^{\star}$, we apply the proximal-gradient homotopy update \eqref{Proximal_lowrank} with $\mu  = 1/m$ and $\Delta_0 \geq \|\bmX^{\star}\|_F$. Let the continuation parameter $\lambda_{t}$ be chosen according to (\ref{Lambda_lowrank}). If
	\begin{equation}\label{sufficentcondition2}
		\sqrt{m} > CK^2\bPa{\gcomp{\calS_{3r} \cap \mathbb{S}^{d^2 - 1}} + \eta},
	\end{equation}
then the update \eqref{Proximal_lowrank} satisfies
	\begin{equation}\label{Interation_convergence_lowrank}
		\fnorm{\bmX_{t+1} - \bmX^{\star}} \leq \rho^{t+1} \Delta_0 + \frac{\xi}{1 - \rho} \frac{\delta}{m}
	\end{equation}
	with probability at least $1 - c\exp\pa{-\eta^2}$. Here
	\begin{align}
		\rho &=  \frac{C'K^2}{\sqrt{m}}\Big(\gcomp{\calS_{3r} \cap \mathbb{S}^{d^2 - 1}} + \eta\Big), \label{ParameterRho} \\
		\xi  &=  C''K (\gcomp{\calS_{3r} \cap \mathbb{S}^{d^2 - 1}} + \eta),  \label{ParameterXi}\\
        \Delta_{t + 1} &= \rho \Delta_t + \frac{1}{m} \xi \delta, \label{updateofdelta}
	\end{align}
and $c, C, C', C''$ are absolute constants.
\end{theorem}

\begin{remark}[Sharpness]
	It follows from \cite[ex. 10.4.4]{vershynin2018HDP} that the Gaussian complexity $\gamma(\calS_r \cap \mathbb{S}^{d^2 - 1})$ can be bounded as $\gcomp{\calS_r \cap \mathbb{S}^{d^2 - 1}} = \gwidth{\calS_r \cap \mathbb{S}^{d^2 - 1}} \lesssim \sqrt{rd}$, where the equality holds because of the symmetry of $\calS_r$. This implies that our sufficient condition \eqref{sufficentcondition2} is sharp up to an absolute constant factor.
\end{remark}
\begin{remark}[Linear convergence rate]
	In the absence of noise, i.e., $\delta = 0$, our analysis shows that if the number of measurements is large enough, then the proximal-gradient homotopy update \eqref{Proximal_lowrank} has a linear convergence rate. Moreover, \eqref{ParameterRho} and \eqref{Interation_convergence_lowrank} indicate that more observations will lead to a smaller $\rho$, and hence a faster convergence speed. In addition, in the presence of noise, \eqref{ParameterRho} and \eqref{Interation_convergence_lowrank} also reveal that more measurements will lead to less estimation error.
\end{remark}

Based on Lemma \ref{Constraint_lowrank} and Theorem \ref{Convergence_lowrank}, we summarize the proximal-gradient homotopy update for low-rank matrices recovery in Algorithm \ref{alg:3}.

\begin{algorithm}[ht]
	\caption{Proximal-gradient homotopy method for low-rank matrices recovery}
	{\bf Input:} Initial value $\Delta_0$, observation matrix $\calA$, measurements $\bmy$, maximum rank $r$, contraction parameters $\rho$ and $\rho_{r,3r}$, noise parameters $\xi$ and $\xi_{r}$, noise level $\delta$, maximum iteration number $t_{\mathrm{max}}$
	
	\begin{algorithmic}
		\STATE Initialize $\bmX_0 = \bmzero$
		\FOR{$t=0, 1, \ldots, t_{\mathrm{max}}-1$}
		\STATE $\lambda_{t} = \frac{1}{\sqrt{r}}\xi_{r}\delta + \frac{m}{ \sqrt{r}}\rho_{r,3r} \Delta_t$
		\STATE $\Xtp = \Prox{\Xt - \frac{1}{m} \calA^T (\calA (\Xt) - \bmY)}{\frac{1}{m} \lambda_{t} \|\cdot\|_*}$
		\STATE $\Delta_{t+1} = \rho\Delta_t + \frac{1}{m} \xi \delta$
		\ENDFOR
	\end{algorithmic} \label{alg:3}
	{\bf Return} $\bmx_{t_{\mathrm{max}}}$
\end{algorithm}


\section{Numerical results} \label{Numerical results}
In this section, we verify our theoretical results  via  a series of numerical simulations. For simplicity, all experiments are performed under the Gaussian measurements, i.e., $\bmA$ has i.i.d. standard Gaussian entries.

\subsection{Sparse Signals Recovery}
In this case, the desired signal $\xreal$ is sparse. We consider both noiseless and noisy cases. In the latter case, $\bmomega \sim \mathcal{N}(\mathbf{0}, \sigma^2\bmI)$ with $\sigma = 0.001$.
We set the ambient dimension $n = 2000$ and the sparsity level $s = 5$. To illustrate time-data tradeoffs, we conduct the simulations under three scenarios $m_1 = 800$, $m_2 = 1300$, $m_3 = 1800$.
\begin{figure}[htbp]
	\centerline{\subfigure[Noiseless case]{\includegraphics[width=1.8in]{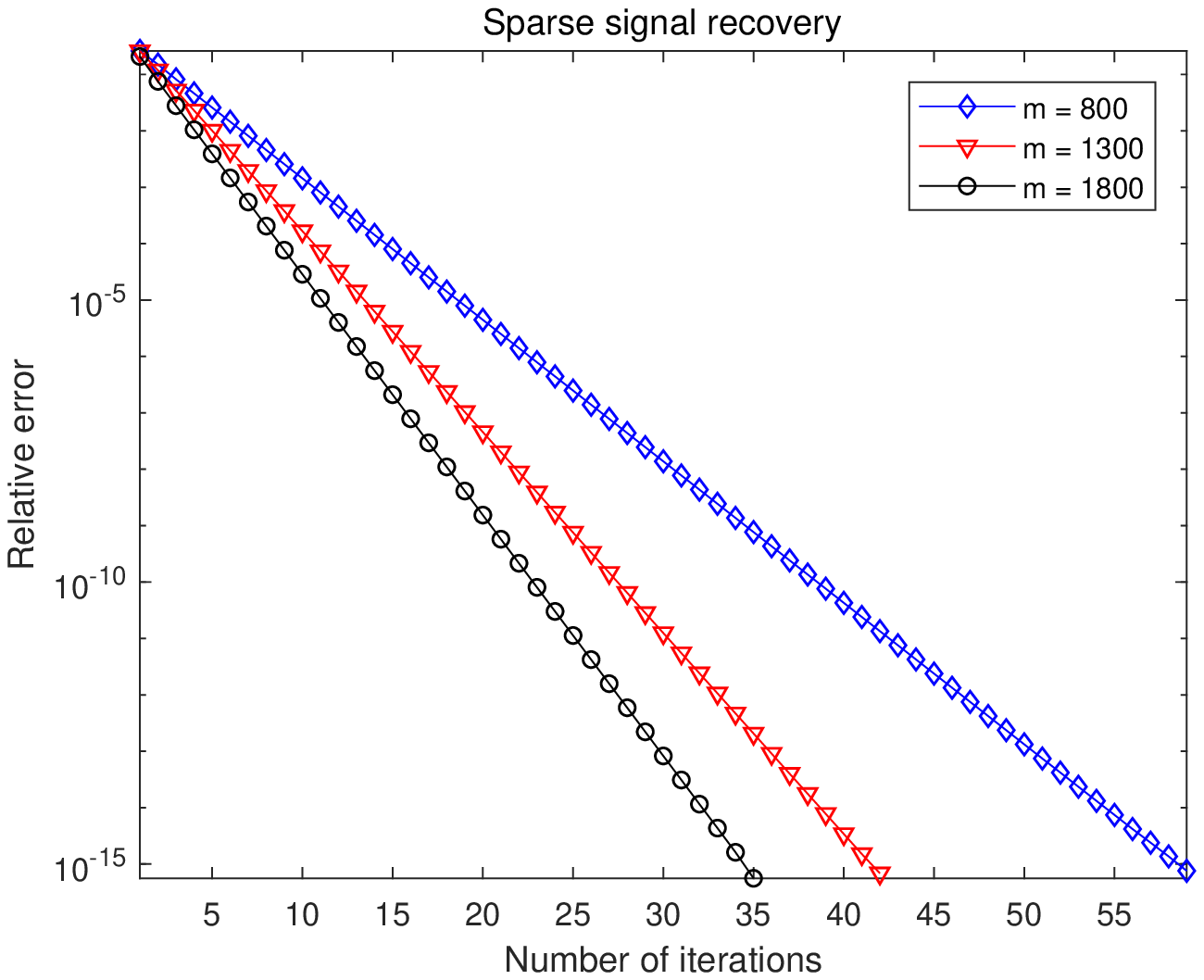}
			\label{PGH sparse mod_noiseless}}
		\hfil
		\subfigure[Noisy case]{\includegraphics[width=1.8in]{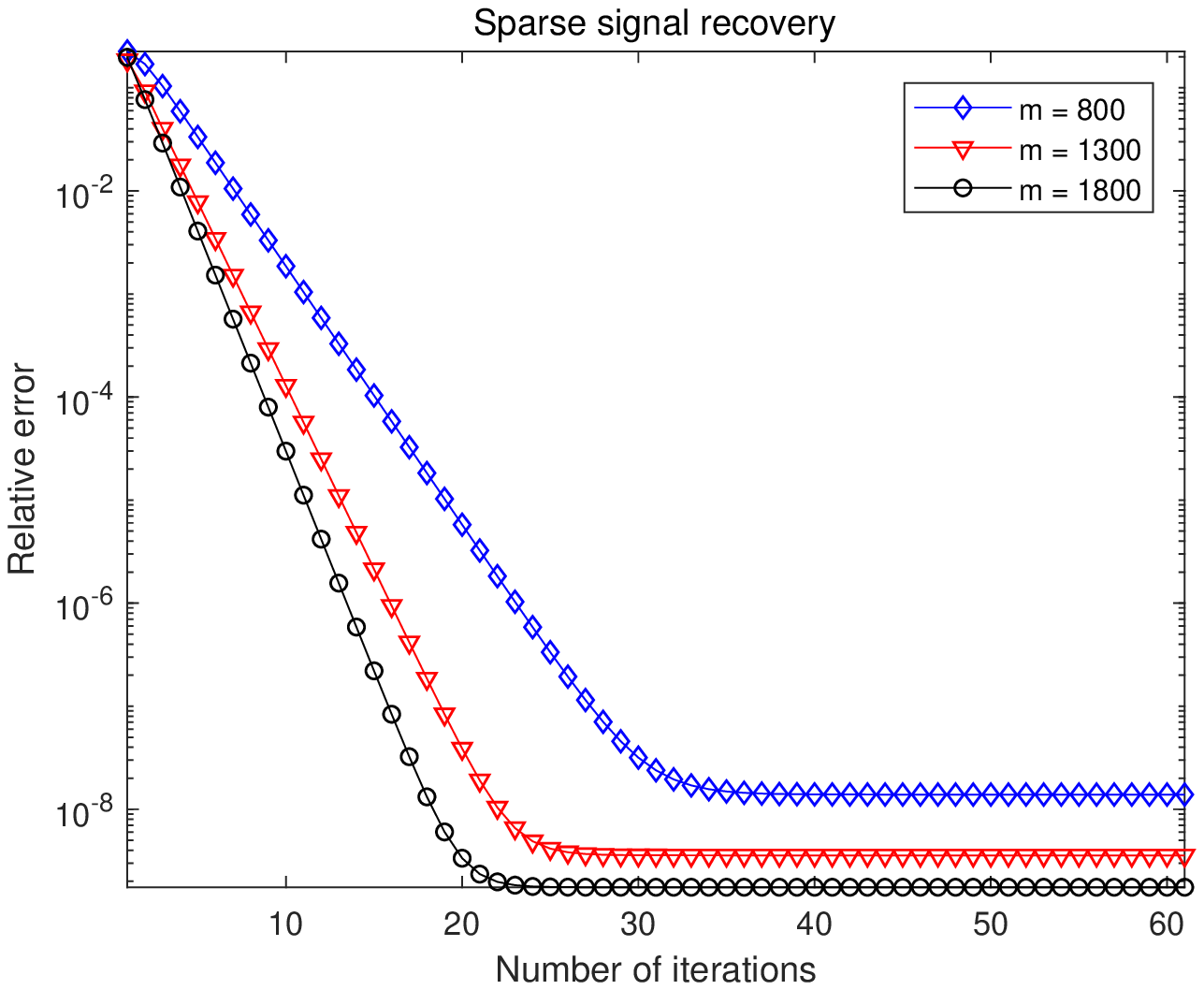}
			\label{PGH sparse mod}}}
	\caption{Convergence results for the sparse signals recovery. }
	\label{PGH_Sparse}
\end{figure}

Figures \ref{PGH sparse mod_noiseless} and \ref{PGH sparse mod} show that the proximal-gradient homotopy update has a linear convergence rate, and that more data would lead to a faster speed. Figure \ref{PGH sparse mod} also illustrates that the estimation error would decrease with the increase of measurements. Both of two diagrams well support our theoretical results in Theorem \ref{Convergence_sparse}.

\subsection{Group Sparse Signals Recovery}
In this case, the targeted signal is group sparse. With the ambient dimension $n = 2500$, we divide the entries of $\bmx \in \RR^n$ into $500$ parts uniformly. We set $s = 5$, which means that there are $5$ non-zero parts in this signal.  We consider both noiseless and noisy cases. In the latter case, $\bmomega \sim \mathcal{N}(\mathbf{0}, \sigma^2\bmI)$ with $\sigma = 0.001$.
To illustrate the time-data tradeoffs, we conduct the simulations under three scenarios $m_1 = 1200$, $m_2 = 1800$, $m_3 = 2400$.
\begin{figure}[htbp]
	\centerline{\subfigure[Noiseless case]{\includegraphics[width=1.85in]{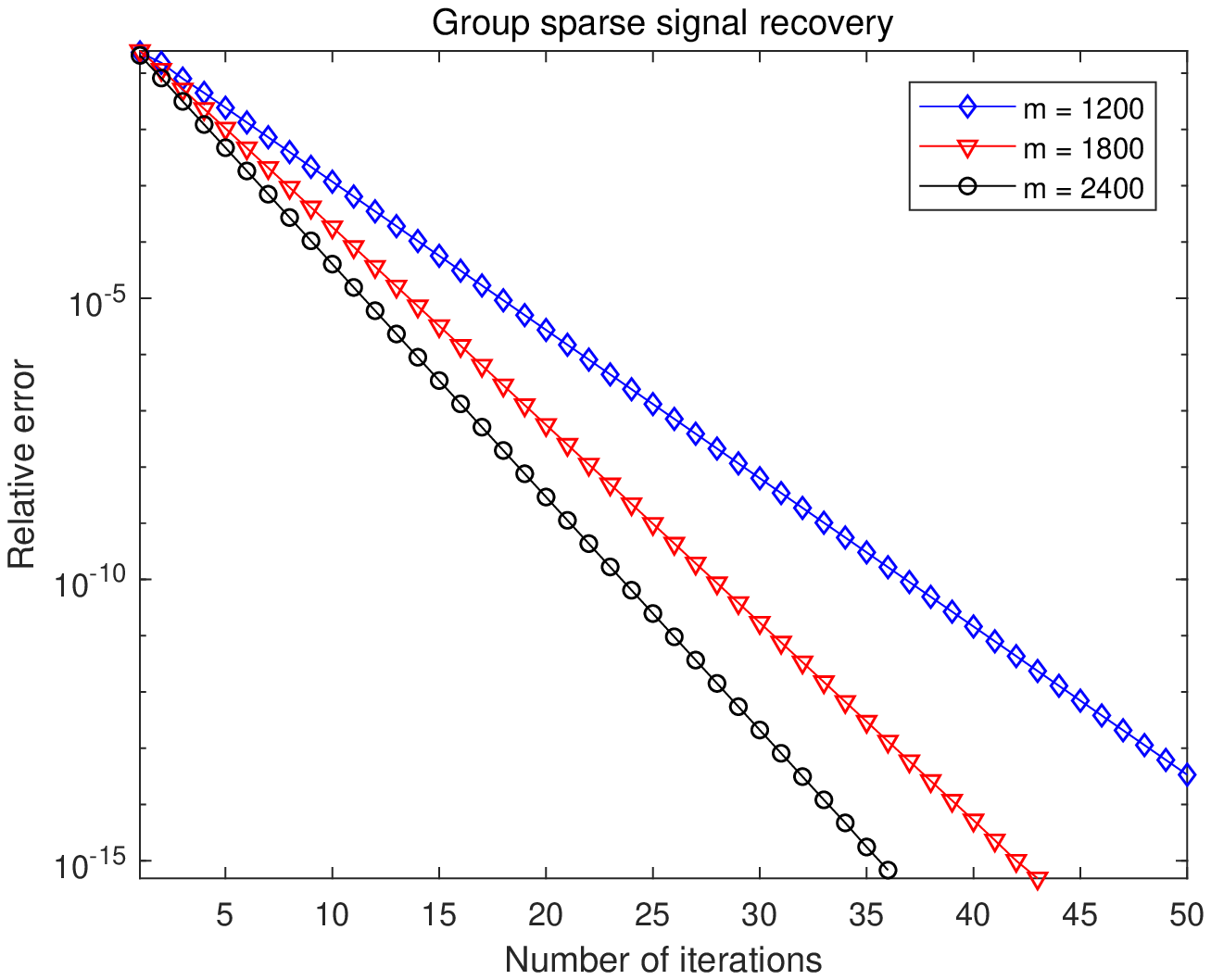}
			\label{PGH Group Mod_Noiseless}}
		\hfil
		\subfigure[Noisy case]{\includegraphics[width=1.85in]{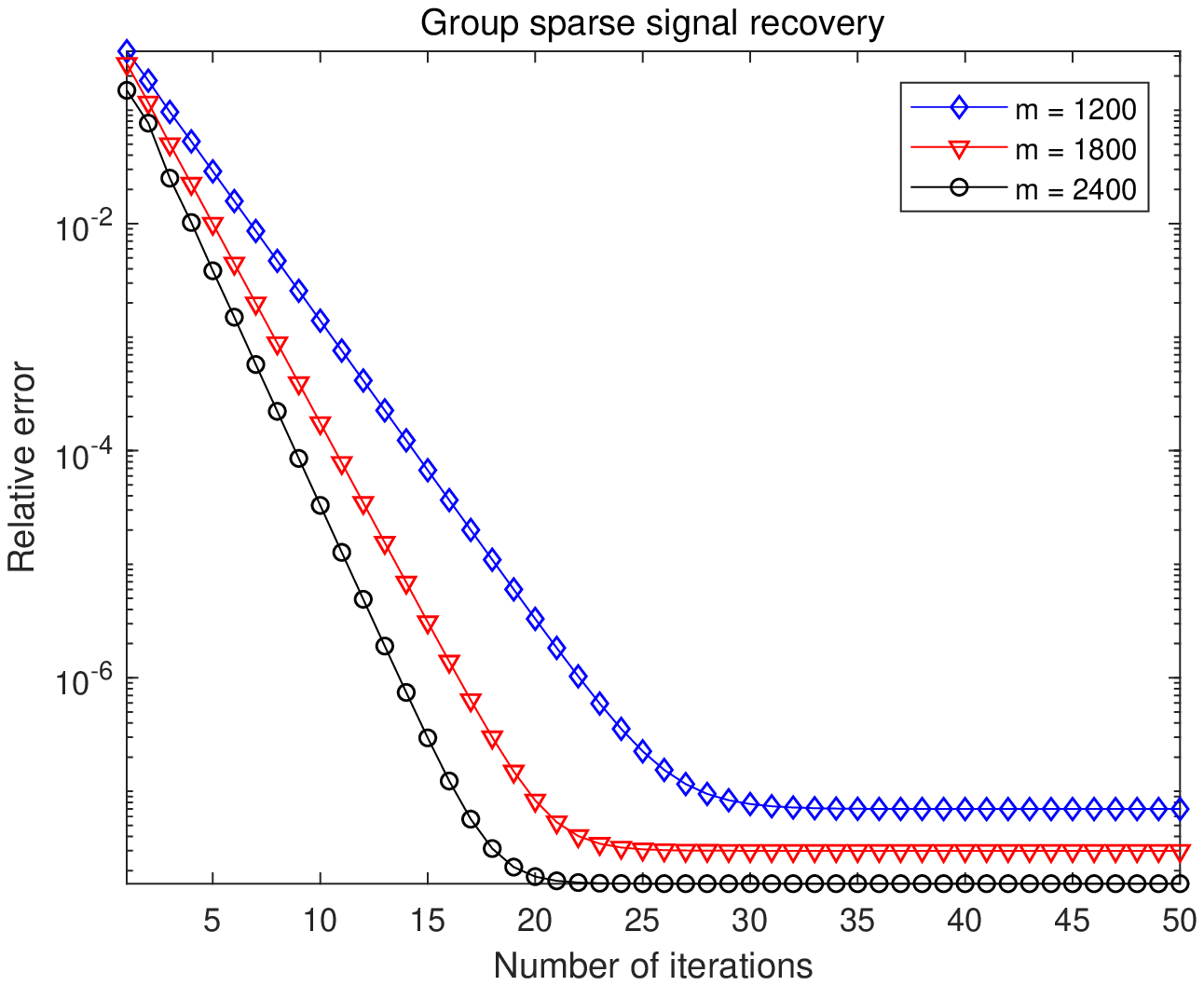}
			\label{PGH Group Mod}}}
	\caption{Convergence results for the group sparse signals recovery. }
	\label{PGH_Group}
\end{figure}

Figures \ref{PGH Group Mod_Noiseless} and \ref{PGH Group Mod} indicate the linear convergence rate of $\ltwonorm{\xt - \xreal} / \ltwonorm{\xreal}$, and reveal that more data would lead to a faster speed. Figure \ref{PGH Group Mod} also shows that the estimation error would decrease with the increase of measurements. The two diagrams also support Remark \ref{RemarkGroup}.

\subsection{Low-Rank Matrices Recovery}
In this case, the desired signal $\Xreal$ is a low-rank matrix. We set the ambient dimension $d \times d = 50 \times 50$ and the rank $r = 2$. $\Xreal$ is generated by the product $\bmU \bmV^T$, where $\bmU, \bmV \in \RR^{50 \times 2}$ are matrices with i.i.d. standard Gaussian entries. We consider both noiseless and noisy cases. In the latter case, $\bmomega \sim \mathcal{N}(\mathbf{0}, \sigma^2\bmI)$ with $\sigma = 0.001$.  To illustrate the time-data tradeoffs, we conduct the simulations under three scenarios $m_1 = 40^2$, $m_2 = 44^2$, $m_3 = 48^2$.
\begin{figure}[htbp]
	\centerline{\subfigure[Noiseless case]{\includegraphics[width=1.85in]{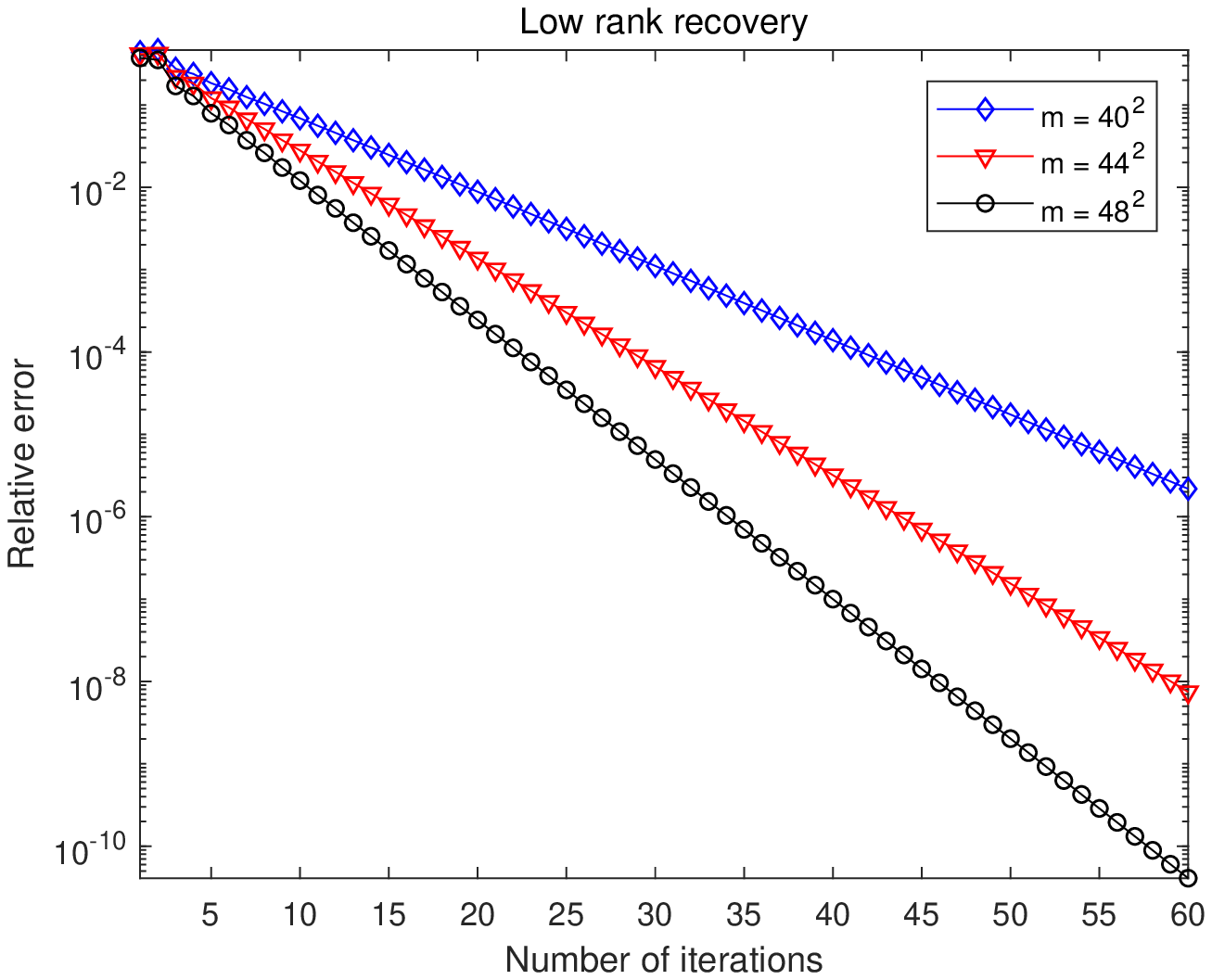}
			\label{PGH Low Mod_Noiseless}}
		\hfil
		\subfigure[Noisy case]{\includegraphics[width=1.85in]{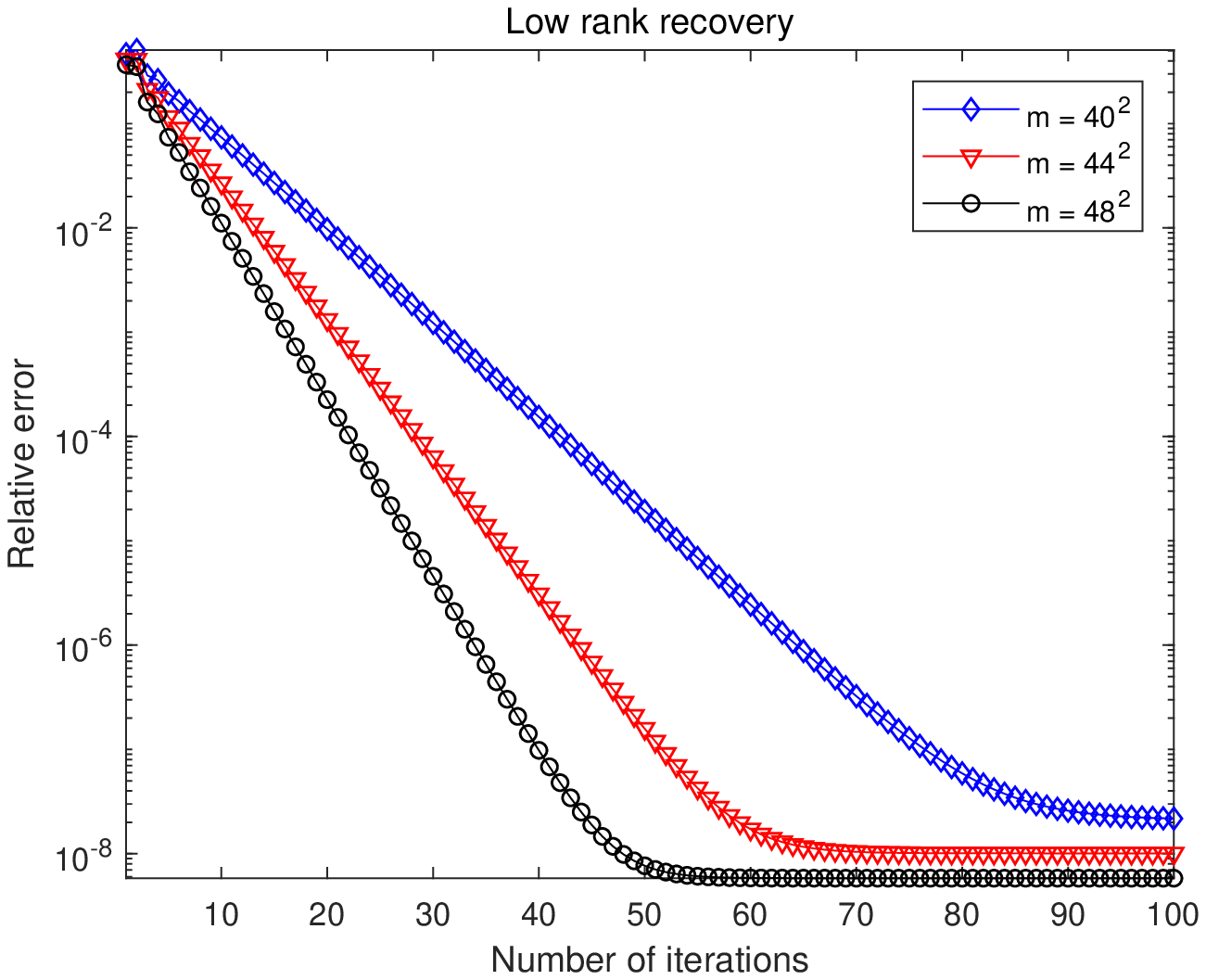}
			\label{PGH Low Mod}}}
	\caption{Convergence results for the low-rank matrices recovery. }
	\label{PGH_Lowrank}
\end{figure}

Figures \ref{PGH Low Mod_Noiseless} and \ref{PGH Low Mod} indicate the linear convergence rate of $\fnorm{\Xt - \Xreal} / \fnorm{\Xreal}$,  and show that more data would lead to a faster speed. Figure \ref{PGH Low Mod} also illustrates the estimation error would decrease with the increase of measurements. These numerical results support our theoretical results in Theorem \ref{Convergence_lowrank}.


\section{Conclusion} \label{Discussion}
In this paper, we have established time-data tradeoffs in solving the penalized linear inverse problem via the proximal-gradient homotopy method. Our strategy is to carefully tune the continuation parameter such that the proximal-gradient homotopy update  belongs to a certain structural constraint set. Our analysis removes the resampling assumption in \cite{Oymak2017FastR}. Our ongoing works is to consider general convex regularization $\Reg{\cdot}$.

\appendices

\section{Proofs of Lemma \ref{Constraint_sparse} and Theorem \ref{Convergence_sparse}}

\begin{proof}[Proof of Lemma \ref{Constraint_sparse}]
	We prove Lemma \ref{Constraint_sparse} by constructing a contradiction. Suppose $\Card(\supp(\xtp) \backslash \supp(\xreal)) \geq s$. This means that $\xtp$ has at least $s$ non-zero entries on the complement of $\supp(\xreal)$.
	
	Since $\xtp$ is an optimal point of the proximal operator \eqref{Proximal_sparse}, there exists a $\ztp \in \partial \lonenorm{\xtp}$  such that
	\begin{equation}\label{Optimal_condition_sparse}
		\lambdat \mu \ztp + \xtp - \xreal - (\bmI - \mu \bmA^{T} \bmA)\htn - \mu \bmA^{T} \bmomega = \bmzero,
	\end{equation}
where $\htn = \bmx_t - \bmx^*$.
	
	We now construct a new vector
    \begin{equation*}
      \bar{\bmz} = \sum_{k = 1}^{s} \sign((\xtp)_{i_k})\bme_{i_k},
    \end{equation*}
  where $\ba{\bme_{i}}_{i = 1}^{n}$ are the standard orthogonal bases and $i_k$ is selected from the set $\ba{i \mid (\xtp)_{i} \neq 0 \text{ and } (\xreal)_{i} = 0}$, which has at least $s$ components by assumption.
  Recall that the subdifferential of the $l_1$-norm is given by
  \begin{equation*}
    \partial\|\bmx\|_1 = \{ \sign(\bmx) + \bmv \},
  \end{equation*}
  where $\bmv_i = 0$, if $i \in \supp(\bmx)$; $|\bmv_i| \leq 1$, otherwise. Then it is not hard to verify that  $\bar{\bmz}$ satisfies the following properties:
   \begin{align*}
     \iprod{\bar{\bmz}}{\ztp} & = s, \\
     \iprod{\bar{\bmz}}{\xtp} & > 0, \\
     \iprod{\bar{\bmz}}{\xreal} & = 0, \\
     \lzeronorm{\bar{\bmz}} & = s, \\
     \ltwonorm{\bar{\bmz}} & = \sqrt{s}.
   \end{align*}

    Taking the inner product with $\bar{\bmz}$ at both sides of the optimal condition \eqref{Optimal_condition_sparse} yields
	\begin{align}
		0   & =  {\lambda_{t}}{\mu} \iprod{\bar{\bmz}}{\ztp} + \iprod{\bar{\bmz}}{\xtp - \xreal} \nonumber \\
		    & ~~~~- \iprod{\bar{\bmz}}{(\bmI - \mu \bmA^T \bmA )\htn} - \mu \iprod{\bar{\bmz}}{\bmA^T \bmomega} \nonumber \\
		    &>  {\lambda_{t}}{\mu} s - \iprod{\bar{\bmz}}{(\bmI - \mu \bmA^T \bmA )\htn} - \mu \iprod{\bar{\bmz}}{\bmA^T \bmomega} \nonumber \\
		    &\geq  {\lambda_{t}}{\mu} s - \sqrt{s}\cdot\rho(\calS_{s} \cap \mathbb{S}^{n - 1},\calS_{2s} \cap \mathbb{S}^{n - 1}) \cdot \Delta_t \nonumber \\
		    & ~~~~-  {\mu}{\sqrt{s}}\cdot \xi(\calS_{s} \cap \mathbb{S}^{n - 1}) \cdot \delta  \nonumber\\
            &  = 0, \nonumber
	\end{align}
   which leads to a contradiction and hence completes the proof. Here, the first inequality holds by noting $\iprod{\bar{\bmz}}{\xtp - \xreal}>0$; the second inequality has used the facts that $\Card(\supp(\htn)) = \Card(\supp(\xt - \xreal)) < 2s$, $\ltwonorm{\htn} \leq \Delta_t$, $\ltwonorm{\bmomega} \leq \delta$, and the definitions \eqref{restricted_singular_values1} and \eqref{restricted_singular_values2}; the last equality follows from the definition of $\lambda_{t}$ (\ref{Lambda_sparse}).
\end{proof}

\begin{proof}[Proof of Theorem \ref{Convergence_sparse}]
For clarity, the proof is divided into two steps.

\textbf{Step 1: Establish deterministic convergence results.} Observe that $\xtp$ is an optimal point of the proximal operator \eqref{Proximal_sparse}, then the optimal condition \eqref{Optimal_condition_sparse} also holds. Taking the inner product with $\htp = \xtp - \xreal$ at both sides of \eqref{Optimal_condition_sparse} yields
\begin{align}\label{Convergence_1}
	\ltwonorm{\htp}^2 &= \iprod{\htp}{(\bmI - \mu \bmA^T \bmA)\htn} - \lambdat \mu \iprod{\htp}{\ztp} \nonumber \\
	&\qquad + \mu \iprod{\htp}{\bmA^{T} \bmomega},
\end{align}
where $\ztp \in \partial \lonenorm{\xtp}$.

We now bound the three terms on the right side of \eqref{Convergence_1} separately. Note that the first term can be bounded as
\begin{align*}
	 &\iprod{\htp}{(\bmI - \mu \bmA^T \bmA)\htn} \\
     &\leq \rho(\calS_{2s} \cap \mathbb{S}^{n - 1},\calS_{2s} \cap \mathbb{S}^{n - 1}) \ltwonorm{\htn}\ltwonorm{\htp}, \\
     & =  \rho_{2s, 2s} \ltwonorm{\htn}\ltwonorm{\htp},
\end{align*}
where we have used the facts that $\Card(\supp(\htn)) < 2s$, $\Card(\supp(\htp)) < 2s$, and the definition \eqref{restricted_singular_values1}.

The second term can be bounded as
\begin{align*}
	- \lambdat \mu \iprod{\htp}{\ztp} &= \lambdat \mu \iprod{\ztp}{\xreal - \xtp} \nonumber \\
	&\leq \lambdat \mu (\lonenorm{\xreal} - \lonenorm{\xtp}) \nonumber \\
	&\leq \lambdat \mu \iprod{\zxstar}{\xreal - \xtp} \nonumber \\
	&\leq \lambdat \mu \ltwonorm{\zxstar}\ltwonorm{\htp} \nonumber \\
	&= \lambdat \mu \sqrt{s}\ltwonorm{\htp},
\end{align*}
where $\zxstar = \uargmin{\bmz \in \partial \lonenorm{\xreal}} \ltwonorm{\bmz}^2$. Here the first two inequalities are due to the definition of the subdifferential and the last inequality has used the Cauchy-Schwartz inequality.

We finally bound the third term as
\begin{align*}
\mu \iprod{\htp}{\bmA^{T} \bmomega} & \leq \mu \cdot \xi(\calS_{2s} \cap \mathbb{S}^{n - 1}) \|\htp\|_2 \|\bmomega\|_2\\
                                    & \leq  \mu \cdot \delta \cdot \xi_{2s} \|\htp\|_2,
\end{align*}
where the first inequality holds because $\Card(\supp(\htp)) < 2s$ and the definition \eqref{restricted_singular_values2}, and the second inequality is due to $\|\bmomega\|_2 \leq \delta$.

Substituting the above three bounds into \eqref{Convergence_1} and noting the definition of $\lambda_{t}$ \eqref{Lambda_sparse} yields
\begin{align} \label{upperbound1}
	&\ltwonorm{\htp} \\ \notag
	&\leq \rho_{2s,2s} \ltwonorm{\htn} + \lambda_{t} \mu \sqrt{s} + \mu \cdot \delta \cdot \xi_{2s}  \\ \notag
	&\leq (\rho_{2s,2s}+\rho_{s,2s}) \cdot \Delta_t + (\xi_{s}+\xi_{2s})\cdot \mu \cdot \delta \\ \notag
    &\leq 2\rho_{2s,2s}\cdot \Delta_t + 2\xi_{2s}\cdot \mu \cdot \delta,
\end{align}
where the second inequality comes from $\ltwonorm{\htn} \leq \Delta_t$ and the last inequality has used the facts $\rho_{s,2s} \leq \rho_{2s,2s}$ and $\xi_{s} \leq \xi_{2s}$.

\textbf{Step 2: Establish high probability convergence results.} To establish high probability convergence results, it suffices to establish the high probability bounds of $\rho_{2s,2s}$ and $\xi_{2s}$. To this end, we have the following two lemmas.
\begin{lemma} \label{Quadratic_deviation}
	For two sets $\calP \subseteq \mathbb{S}^{n - 1}$ and $\calQ \subseteq \mathbb{S}^{n - 1}$, if the number of measurements satisfies
	\begin{equation}
		\sqrt{m} \geq CK^2 (\gcomp{\calP} + \gcomp{\calQ} + 2\eta), \label{sufficientcondition3}
	\end{equation}
	then the event
		\begin{equation*}
			\usup{\bmu \in \calP, \bmv \in \calQ} \iprod{\bmv}{(\bmI - \frac{\bmA^{T}\bmA}{m})\bmu} \leq C'K^2 \frac{\gcomp{\calP} + \gcomp{\calQ} + 2\eta}{\sqrt{m}}
		\end{equation*}
	holds with probability at least $1 - 2\exp(-\eta^2)$.
\end{lemma}
\begin{proof}
  See Appendix \ref{ProofofQuadratic_deviation}.
\end{proof}

\begin{lemma} \cite[Lemma 4]{ChenJ2019StableR}
\label{lm: upper bound of general ip}
Let $\bmA$ be a matrix whose rows $\bmA_i$ are independent centered isotropic sub-Gaussian vectors with $\max_{i} \|\bmA_i\|_{\psi_2} \leq K$, and $\bmw $ be any fixed vector. Let $\calT$ be any bounded subset $\RR^n$. Then, for any $\eta\geq 0$, the event
\begin{align*}
	\sup_{\bmu \in \calT} \langle{\bmA\bmu}, {\bmw}\rangle \leq CK \|\bmw\|_2\big[ \gamma(\calT) + \eta\cdot\rad\calT \big]
\end{align*}
holds with probability at least $1-\exp\{-\eta^2\}$.
\end{lemma}

It then follows from Lemma \ref{Quadratic_deviation} with the number of measurements satisfying \eqref{sufficentcondition1} and Lemma \ref{lm: upper bound of general ip} that the event
\begin{equation*}
  \rho_{2s,2s}   \leq 2CK^2 \frac{\gcomp{\calS_{2s} \cap \mathbb{S}^{n - 1}} + \eta}{\sqrt{m}}
\end{equation*}
holds with probability at least $1 - 2\exp(-\eta^2)$, and the event
\begin{equation*}
  \xi_{2s}   \leq CK (\gcomp{\calS_{2s} \cap \mathbb{S}^{n - 1}} + \eta)
\end{equation*}
holds with probability at least $1 - \exp(-\eta^2)$.

Substituting the above two bounds into \eqref{upperbound1} and taking the union bound yields the event
\begin{align} \label{upperbound2}
	&\ltwonorm{\htp} \\ \notag
    &\leq \underbrace{4CK^2 \frac{\gcomp{\calS_{2s} \cap \mathbb{S}^{n - 1}} + \eta}{\sqrt{m}}}_{:=\rho}\cdot \Delta_t \\ \notag
    & ~~~~~~+ \underbrace{2CK (\gcomp{\calS_{2s} \cap \mathbb{S}^{n - 1}} + \eta)}_{:=\xi}\cdot \mu \cdot \delta \\ \notag
    & = \rho  \cdot \Delta_t + {\xi}\cdot \mu \cdot \delta = \Delta_{t+1} \\ \notag
    & \leq \rho^{t+1} \cdot \Delta_0 + \frac{1}{1-\rho}{\xi}\cdot \mu \cdot \delta
\end{align}
holds with probability at least $1 - c\exp(-\eta^2)$. The last inequality holds because the condition \eqref{sufficentcondition1} ensures $\rho < 1$. Thus we complete the proof.
\end{proof}

\section{Proofs of Lemma \ref{Constraint_lowrank} and Theorem \ref{Convergence_lowrank}}

\begin{proof}[Proof of Lemma \ref{Constraint_lowrank}]
   We prove Lemma \ref{Constraint_lowrank} also by constructing a contradiction.
   Suppose $\rank(\Xtp) \geq 2r$. Since $\Xtp$ is the optimal point of the proximal operator (\ref{Proximal_lowrank}), there exists a $\Ztp \in \partial \nnorm{\Xtp}$ such that
\begin{equation}
	\lambdat \mu \Ztp + \Xtp - \Xreal - (\calI - \mu \calA^{T} \calA)(\Htn) - \mu \calA^{T} (\bmomega) = \bmzero. \label{Optimal_condition_lowrank}
\end{equation}

We define a new matrix $\bar{\bmZ} = \sum_{i = 1}^{r} \bmu'_i (\bmv'_i)^T$, where $\ba{\bmu_{i}'}_{i = 1}^{r}$ and $\ba{\bmv_{i}'}_{i = 1}^{r}$ are two sets of orthonormal vectors constructed as follows.

Let $\Xtp = \bmU_{t + 1} \Sigmatp \bmV_{t + 1}^T$ and $\Xreal = \bmU_{\star} \Sigmastar \bmV_{\star}^T$ be the singular value decompositions of $\Xtp$ and $\Xreal$ respectively. Let $\ba{\bmu_{i}^{t+1}}_{i = 1}^{2r}$ be the first $2r$ columns of $\bmU_{t + 1}$ and $\ba{\bmu_{i}^{\star}}_{i = 1}^{r}$ be the first $r$ columns of $\bmU_{\star}$. Then we have
\begin{align*}
	& \dim[\Span \ba{\utp_1, \ldots, \utp_{2r}} \cap (\Span \ba{\ureal_1, \ldots, \ureal_{r}})^{\perp}] \\
	& \geq 2r + (d - r) - d \\
	& = r,
\end{align*}
where we have used the facts that
\begin{equation}
	\dim(\calU \cap \calW) = \dim(\calU) + \dim(\calW) - \dim(\calU + \calW) \nonumber
\end{equation}
for two subspaces $\calU$ and $\calW$ \cite[pp. 47]{Axler2015LinearA}, and $\dim(\Span \ba{\utp_1, \cdots, \utp_{2r}} \cup (\Span \ba{\ureal_1, \cdots, \ureal_{r}})^{\perp}) \leq d$. Thus we can find a set of orthonormal vectors $\ba{\bmu_{i}'}_{i = 1}^{r}$ from the intersection of $\Span \ba{\utp_1, \cdots, \utp_{2r}}$ and $(\Span \ba{\ureal_1, \cdots, \ureal_{r}})^{\perp}$.  Moreover, there exist $\{\gamma_{ij}\}$ such that
\begin{equation*}
	\bmu'_i = \sum_{j = 1}^{2r} \gamma_{ij} \utp_j = \bmU_{t + 1}^{2r} \bmgamma_{i}, ~~~~i = 1, \cdots, r,
\end{equation*}
where $\bmU_{t + 1}^{2r} = [\bmu_{1}^{t + 1}, \cdots, \bmu_{2r}^{t + 1}]$ and $\bmgamma_i = (\gamma_{i1}, \cdots, \gamma_{i2r})^T$. It is not hard to verify $\{\bmgamma_{i}\}$ are orthonormal vectors. To see this, observe that
\begin{equation*}
	\iprod{\bmu'_{i}}{\bmu'_{j}} = \bmgamma_{i}^T (\bmU_{t + 1}^{2r})^T \bmU_{t + 1}^{2r} \bmgamma_{j} = \iprod{\bmgamma_{i}}{\bmgamma_{j}}.
\end{equation*}
If $i \neq j$, then we have $\iprod{\bmgamma_{i}}{\bmgamma_{j}} = 0$; if $i = j$, then we have $\ltwonorm{\bmgamma_{i}}^2 = 1$, for $i = 1, \cdots, r$. Here we have used the facts that $\ba{\bmu_{i}'}_{i = 1}^{r}$ are a set of orthonormal vectors and $(\bmU_{t + 1}^{2r})^T \bmU_{t + 1}^{2r} = \bmI_{2r}$.

We then construct $\ba{\bmv_{i}'}_{i = 1}^{r}$ as follows
\begin{equation*}
	\bmv'_i = \bmV_{t + 1}^{2r} \bmgamma_{i}, ~~~~i = 1, \cdots, r,
\end{equation*}
where $\bmV_{t + 1}^{2r} = [\bmv_{1}^{t + 1}, \cdots, \bmv_{2r}^{t + 1}]$ is composed of the first $2r$ columns of $\bmV_{t + 1}$. Clearly, $\ba{\bmv_{i}'}_{i = 1}^{r}$ are also a set of orthonormal vectors by noting that
\begin{equation*}
	\iprod{\bmv_{i}'}{\bmv_{j}'} = \bmgamma_{i}^T (\bmV_{t + 1}^{2r})^T \bmV_{t + 1}^{2r} \bmgamma_{j} = \iprod{\bmgamma_{i}}{\bmgamma_{j}},
\end{equation*}
and $\{\bmgamma_{i}\}$ are orthonormal vectors.

We now verify that $\bar{\bmZ}$ satisfies the following properties:
\begin{align}
	\iprod{\bar{\bmZ}}{\Ztp} &= r \label{Property_lowrank1} \\
	\iprod{\bar{\bmZ}}{\Xtp} &> 0 \label{Property_lowrank2} \\
	\iprod{\bar{\bmZ}}{\Xreal} &= 0 \label{Property_lowrank3} \\
	\rank(\bar{\bmZ}) &= r \label{Property_lowrank4} \\
	\fnorm{\bar{\bmZ}} &= \sqrt{r}.  \label{Property_lowrank5}
\end{align}

Let $k$ be the rank of $\Xtp$ ($k \geq 2r$ by assumption). Recall that the subdifferential of the nuclear norm at $\Xtp$ is given by \cite[pp. 40]{Watson1992CharacterizationOfSubd}
\begin{equation}
	\partial \|\bmX_{t+1}\|_* = \ba{\bmU_{t+1}^{k} (\bmV_{t+1}^k)^T + \bmW \mid \bmW \in T_{\bmX_{t+1}}^{\perp}, ~\norm{\bmW} \leq 1}, \label{SubdifferentialLowrank}
\end{equation}
where $T_{\bmX_{t+1}} = \ba{\bmU_{t+1}^{k} \bmA^T + \bmB (\bmV_{t+1}^k)^T \mid \bmA \in \RR^{d \times k}, \bmB \in \RR^{d \times k}}$ and $\norm{\bmW}$ is the spectral norm of $\bmW$.

Note that $\bar{\bmZ}$ can be reformulated as
\begin{equation*}
	\begin{split}
		\bar{\bmZ} &= \sum_{i = 1}^{r} \bmu'_i (\bmv'_i)^T = \sum_{i = 1}^{r} \bmU_{t + 1}^{2r} \bmgamma_{i} \bmgamma_{i}^T (\bmV_{t + 1}^{2r})^T \\
		&= \bmU_{t + 1}^{2r} (\sum_{i = 1}^{r} \bmgamma_{i} \bmgamma_{i}^T) (\bmV_{t + 1}^{2r})^T.
	\end{split}
\end{equation*}

Thus (\ref{Property_lowrank1}) follows from
\begin{equation*}
	\begin{split}
		\iprod{\bar{\bmZ}}{\Ztp} &= \iprod{\bmU_{t + 1}^{2r} (\sum_{i = 1}^{r} \bmgamma_{i} \bmgamma_{i}^T) (\bmV_{t + 1}^{2r})^T}{\bmU_{t + 1}^{k} (\bmV_{t + 1}^{k})^T+\bmW} \\
		&= \tr\Big(\sum_{i = 1}^{r} \bmgamma_{i} \bmgamma_{i}^T\Big) \\
		&= \sum_{i = 1}^{r} \ltwonorm{\bmgamma_{i}}^2 = r,
	\end{split}
\end{equation*}
where we have used the fact that $\bmW$ is orthogonal to $\bar{\bmZ}$.

(\ref{Property_lowrank2}) holds because
\begin{equation*}
	\begin{split}
		\iprod{\bar{\bmZ}}{\Xtp} &= \iprod{\bmU_{t + 1}^{2r} (\sum_{i = 1}^{r} \bmgamma_{i} \bmgamma_{i}^T) (\bmV_{t + 1}^{2r})^T}{\bmU_{t + 1} \bmSigma_{t + 1} (\bmV_{t + 1})^T} \\
		&= \tr\Big((\sum_{i = 1}^{r} \bmgamma_{i} \bmgamma_{i}^T) \bmSigma_{t + 1}^{2r}\Big) \\
		&= \sum_{i = 1}^{r} \bmgamma_{i}^T \bmSigma_{t + 1}^{2r} \bmgamma_{i} > 0,
	\end{split}
\end{equation*}
where $\bmSigma_{t + 1}^{2r}$  is the submatrix formed from rows $1$ through $2r$ and columns $1$ through $2r$ of $\bmSigma_{t + 1}$. Here the second equality has used the assumption $\rank(\Xtp) \geq 2r$ and the last inequality holds because $\bmSigma_{t + 1}^{2r}$ is positive definite.

\eqref{Property_lowrank3} is due to the orthogonality between $\ba{\bmu_{i}'}_{i = 1}^{r}$ and $\ba{\bmu_{i}^{\star}}_{i = 1}^{r}$.

(\ref{Property_lowrank4}) and  (\ref{Property_lowrank5}) easily follow from the construction of $\ba{\bmu_{i}'}_{i = 1}^{r}$ and $\ba{\bmv_{i}'}_{i = 1}^{r}$.

Finally, taking the inner product with $\bar{\bmZ}$ at both sides of the optimal condition \eqref{Optimal_condition_lowrank} yields
\begin{align*}
	0 &=  \lambda_{t} \mu \iprod{\Ztp}{\bar{\bmZ}} + \iprod{\Xtp - \Xreal}{\bar{\bmZ}} \nonumber \\
	&\qquad - \iprod{(\calI - \mu \calA^{T} \calA)(\Htn)}{\bar{\bmZ}} - \mu \iprod{\bar{\bmZ}}{\calA^{T} (\bmomega)} \\
	&> \lambda_{t} \mu r - \iprod{(\calI - \mu \calA^{T} \calA)(\Htn)}{\bar{\bmZ}} - \mu \iprod{\bar{\bmZ}}{\calA^{T} (\bmomega)} \\
	&\geq \lambda_{t} \mu r - \sqrt{r}\rho(\calS_{r} \cap \mathbb{S}^{d^2 - 1},\calS_{3r} \cap \mathbb{S}^{d^2 - 1}) \cdot \Delta_t \nonumber \\
	&\qquad - \mu \sqrt{r}\xi(\calS_{r} \cap \mathbb{S}^{d^2 - 1}) \cdot \delta \nonumber \\
	&= 0,
\end{align*}
which leads to a contradiction and hence completes the proof. Here, the first inequality holds by noting $\iprod{\bar{\bmZ}}{\Xtp - \Xreal}>0$; the second inequality has used the facts that $\rank(\Xt - \Xreal) < 3r$, $\fnorm{\Htn} \leq \Delta_t$, and $\ltwonorm{\bmomega} \leq \delta$; the last equality follows from the definition of $\lambda_{t}$ (\ref{Lambda_lowrank}).

\end{proof}

\begin{proof}[Proof of Theorem \ref{Convergence_lowrank}]
	For clarity, the proof is similarly divided into two steps.
	
	\textbf{Step 1: Establish deterministic convergence results.} Observe that $\Xtp$ is an optimal point of the proximal operator \eqref{Proximal_lowrank}, then the optimal condition \eqref{Optimal_condition_lowrank} also holds. Taking the inner product with $\Htp = \Xtp - \Xreal$ at the both sides of \eqref{Optimal_condition_lowrank} yields
\begin{align} \label{Convergence_2}
		\fnorm{\Htp}^2 &= \iprod{\Htp}{(\calI - \mu \calA^{T} \calA)(\Htn)} - \lambdat \mu \iprod{\Htp}{\Ztp} \nonumber \\
		&\qquad + \mu \iprod{\Xtp - \Xreal}{\calA^{T} (\bmomega)},
\end{align}
where $\Ztp \in \partial \nnorm{\Xtp}$.

We now bound the three terms on the right side of \eqref{Convergence_2} separately. Note that the first item could be bounded as
\begin{align*}
	&\iprod{\Htp}{(\calI - \mu \calA^{T} \calA)(\Htn)} \\
	&\leq \rho(\calS_{3r} \cap \mathbb{S}^{d^2 - 1}, \calS_{3r} \cap \mathbb{S}^{d^2 - 1}) \fnorm{\Htn}\fnorm{\Htp} \\
	&= \rho_{3r,3r} \fnorm{\Htn}\fnorm{\Htp},
\end{align*}
where we have used the facts that $\rank(\Xt - \Xreal) < 3r$, $\rank(\Xtp - \Xreal) < 3r$ and the definition \eqref{restricted_singular_values1}.

The second item could be bounded as
\begin{align*}
	- \lambdat \mu \iprod{\Htp}{\Ztp} &= \lambdat \mu \iprod{\Ztp}{\Xreal - \Xtp} \nonumber \\
	&\leq \lambdat \mu (\nnorm{\Xreal} - \nnorm{\Xtp}) \nonumber \\
	&\leq \lambdat \mu \iprod{\bmZ_{\bmX^{\star}}}{\Xreal - \Xtp} \nonumber \\
	&\leq \lambdat \mu \fnorm{\bmZ_{\bmX^{\star}}}\fnorm{\Htp} \nonumber \\
	&= \lambdat \mu \sqrt{r}\fnorm{\Htp},
\end{align*}
where $\bmZ_{\bmX^{\star}} = \uargmin{\bmZ \in \partial \nnorm{\Xreal}} \fnorm{\bmZ}^2$. Here the first two inequalities are due to the definition of the subdifferential and the last inequality has used the Cauchy-Schwartz inequality.

We finally bound the third term as
\begin{align*}
	\mu \iprod{\Xtp - \Xreal}{\calA^{T} (\bmomega)} &\leq \mu \cdot \xi(\calS_{3r} \cap \mathbb{S}^{d^2 - 1}) \fnorm{\Htp} \ltwonorm{\bmomega} \\
	&\leq \mu \cdot \delta \cdot \xi_{3r} \fnorm{\Htp} ,
\end{align*}
where the first inequality holds because  $\rank(\Xtp - \Xreal) < 3r$ and the definition \eqref{restricted_singular_values2}, and the second inequality is due to $\ltwonorm{\bmomega} \leq \delta$.

Substituting the above three bounds into \eqref{Convergence_2} and noting the definition of $\lambda_{t}$ \eqref{Lambda_lowrank} yields
\begin{align}
	&\fnorm{\Htp} \nonumber \\
	&\leq \rho_{3r,3r} \fnorm{\Htn} + \lambda_{t} \mu \sqrt{r} + \mu \cdot \delta \cdot \xi_{3r} \nonumber \\
	&\leq (\rho_{3r,3r} + \rho_{r,3r}) \cdot \Delta_t + (\xi_{r} + \xi_{3r}) \cdot \mu \cdot \delta \nonumber \\
	&\leq 2 \rho_{3r,3r} \cdot \Delta_t + 2 \xi_{3r} \cdot \mu \cdot \delta, \label{upperbound3}
\end{align}
where the second inequality comes from $\fnorm{\Htn} \leq \Delta_t$ and the definition of \eqref{Lambda_lowrank}, and the last inequality has used the facts $\rho_{r,3r} \leq \rho_{3r,3r}$, $\xi_{r} \leq \xi_{3r}$.

\textbf{Step 2: Establish high probability convergence results.} To establish high probability convergence results, it suffices to establish the high probability bounds of $\rho_{3r,3r}$ and $\xi_{3r}$.

It then follows from Lemma 3 with the number of measurements satisfying \eqref{sufficentcondition2} and Lemma \ref{lm: upper bound of general ip} that the event
\begin{equation*}
	\rho_{3r,3r} \leq 2CK^2 \frac{\gcomp{\calS_{3r} \cap \mathbb{S}^{d^2 - 1}} + \eta}{\sqrt{m}}
\end{equation*}
holds with probability at least $1 - 2\exp(-\eta^2)$, and the event
\begin{equation*}
	\xi_{3r}  \leq CK (\gcomp{\calS_{3r} \cap \mathbb{S}^{d^2 - 1}} + \eta)
\end{equation*}
holds with probability at least $1 - \exp(-\eta^2)$.

Substituting the above two bounds into \eqref{upperbound3} and taking the union bound yields the event
\begin{align} \label{upperbound4}
	&\fnorm{\Htp} \\ \notag
	&\leq \underbrace{4CK^2 \frac{\gcomp{\calS_{3r} \cap \mathbb{S}^{d^2 - 1}} + \eta}{\sqrt{m}}}_{:=\rho}\cdot \Delta_t \\ \notag
	& ~~~~~~+ \underbrace{2CK (\gcomp{\calS_{3r} \cap \mathbb{S}^{d^2 - 1}} + \eta)}_{:=\xi}\cdot \mu \cdot \delta \\ \notag
	& = \rho  \cdot \Delta_t + {\xi}\cdot \mu \cdot \delta = \Delta_{t+1} \\ \notag
	& \leq \rho^{t+1} \cdot \Delta_0 + \frac{1}{1-\rho}{\xi}\cdot \mu \cdot \delta
\end{align}
holds with probability at least $1 - c\exp(-\eta^2)$. The last inequality holds because the condition \eqref{sufficentcondition2} ensures $\rho < 1$. Thus we complete the proof.

\end{proof}

\section{Proof of Lemma \ref{Quadratic_deviation}} \label{ProofofQuadratic_deviation}
 To prove Lemma \ref{Quadratic_deviation}, we require the following matrix deviation inequality for sub-Gaussian matrices.

\begin{fact} \cite[Theorem 3]{Liaw2017ASimpleT} \label{Derivation}
	Let $\bmA$ be a sub-Gaussian matrix with independent centered isotropic rows $\bmA_i$ satisfying $\|\bmA_i\|_{\psi_2} \leq K$. Let $\calT$ be a bounded subset of $\RR^n$. Then for any $\eta \geq 0$, the event
	\begin{align}
		\usup{\bmu \in \calT} \abs{\ltwonorm{\bmA \bmu} - \sqrt{m} \ltwonorm{\bmu}} \leq CK^2 [\gcomp{\calT} + \eta \rad{\calT}]
	\end{align}
	holds with probability at least $1 - \exp(-\eta^2)$, where $\rad{\calT}:= \sup_{\bmx \in \calT}\|\bmx\|_2$ denotes the radius of $\calT$.
\end{fact}

\begin{proof}[Proof of Lemma \ref{Quadratic_deviation}]
	Define the following two sets
	\begin{align*}
	\calT_{-} = \calP - \calQ \quad \text{and} \quad \calT_{+} = \calP + \calQ.
	\end{align*}
	Clearly, we have
	\begin{align*}
	\usup{\bmu \in \calT_{-}} \ltwonorm{\bmu} \leq 2,~~~~\usup{\bmu \in \calT_{+}} \ltwonorm{\bmu} \leq 2,
	\end{align*}
	and
	\begin{align*}
	\gcomp{\calT_+} &= \EE [\usup{\bmu \in \calP, \bmv \in \calQ } \abs{\iprod{\bmg}{\bmu + \bmv}}]  \\
	&\leq \EE [\usup{\bmu \in \calP} \abs{\iprod{\bmg}{\bmu}} +\usup{\bmv \in \calQ} \abs{\iprod{\bmg}{\bmv}}]  \\
	&\leq \gcomp{\calP} + \gcomp{\calQ} :=  \gamma_1 + \gamma_2,\\
	\gcomp{\calT_-} &\leq \gcomp{\calP} + \gcomp{\calQ} :=  \gamma_1 + \gamma_2  .
	\end{align*}	
	
  Observe that the inner product to be bounded can be reformulated as
	\begin{align}\label{Classification_discussion}
	&\iprod{\bmv}{(\bmI - \frac{\bmA^T\bmA}{m})\bmu} \nonumber \\
	&= \frac{1}{4} \Big[\iprod{\bmu + \bmv}{(\bmI - \frac{\bmA^T\bmA}{m})(\bmu + \bmv)} \nonumber \\
	&\qquad - \iprod{\bmu - \bmv}{(\bmI - \frac{\bmA^T\bmA}{m})(\bmu - \bmv)}\Big] \nonumber \\
	&= \frac{1}{4} \Big[\ltwonorm{\bmu + \bmv}^2 - \frac{1}{m}\ltwonorm{\bmA(\bmu + \bmv)}^2 \nonumber \\
	&\qquad - \ltwonorm{\bmu - \bmv}^2 + \frac{1}{m}\ltwonorm{\bmA(\bmu - \bmv)}^2\Big].
	\end{align}
 So it suffices to establish the low bound of $\ltwonorm{\bmA(\bmu + \bmv)}^2/m$ and the upper bound of $\ltwonorm{\bmA(\bmu - \bmv)}^2/m$. It follows from Fact \ref{Derivation} that the two events
	\begin{align}
	\frac{\ltwonorm{\bmA(\bmu + \bmv)}^2}{m} &\geq \Big(\max\ba{\ltwonorm{\bmu + \bmv} - CK^2\frac{\gamma_1 + \gamma_2 + 2\eta}{\sqrt{m}},0} \Big)^2 \label{Relative_size}
	\end{align}
	and
	\begin{align}
	\frac{\ltwonorm{\bmA(\bmu - \bmv)}^2}{m} &\leq \Big(\ltwonorm{\bmu - \bmv} + CK^2\frac{\gamma_1 + \gamma_2 + 2\eta}{\sqrt{m}}\Big)^2
	\end{align}
	hold simultaneously with probability at least $1 - 2\exp(-\eta^2)$.

Thus \eqref{Classification_discussion} can be bounded as: When $\ltwonorm{\bmu + \bmv} \geq CK^2 \frac{\gamma_1 + \gamma_2 + 2\eta}{\sqrt{m}}$, we have
\begin{align*}
	& \iprod{\bmv}{(\bmI - \frac{\bmA^T\bmA}{m})\bmu} \nonumber \\
    & \leq  \frac{1}{4} \Big[\ltwonorm{\bmu + \bmv}^2 - \Big(\ltwonorm{\bmu + \bmv} - CK^2\frac{\gamma_1 + \gamma_2 + 2\eta}{\sqrt{m}}\Big)^2 \nonumber \\
	&\qquad - \ltwonorm{\bmu - \bmv}^2 + \Big(\ltwonorm{\bmu - \bmv} + CK^2\frac{\gamma_1 + \gamma_2 + 2\eta}{\sqrt{m}}\Big)^2 \Big] \\
	& = \frac{1}{2} CK^2 \frac{\gamma_1 + \gamma_2 + 2\eta}{\sqrt{m}} (\ltwonorm{\bmu + \bmv} + \ltwonorm{\bmu - \bmv}) \nonumber \\
	& \leq \sqrt{2} CK^2 \frac{\gamma_1 + \gamma_2 + 2\eta}{\sqrt{m}}\\
    & = C'K^2 \frac{\gamma_1 + \gamma_2 + 2\eta}{\sqrt{m}},
\end{align*}
where the second inequality is due to  $(\ltwonorm{\bmu + \bmv} + \ltwonorm{\bmu - \bmv})^2 \leq 2(\ltwonorm{\bmu + \bmv}^2 + \ltwonorm{\bmu - \bmv}^2) \leq 8$.

If $\ltwonorm{\bmu + \bmv} < CK^2 \frac{\gamma_1 + \gamma_2 + 2\eta}{\sqrt{m}}$, then we can derive
\begin{align*}
	& \iprod{\bmv}{(\bmI - \frac{\bmA^T\bmA}{m})\bmu} \nonumber \\
    & \leq  \frac{1}{4} \Big[\ltwonorm{\bmu + \bmv}^2 - \ltwonorm{\bmu - \bmv}^2 \nonumber \\
	&\qquad  + \Big(\ltwonorm{\bmu - \bmv} + CK^2\frac{\gamma_1 + \gamma_2 + 2\eta}{\sqrt{m}}\Big)^2 \Big] \\
	&\leq \frac{1}{2} \Big(CK^2 \frac{\gamma_1 + \gamma_2 + 2\eta}{\sqrt{m}}\Big)^2 + \frac{1}{2} CK^2 \frac{\gamma_1 + \gamma_2 + 2\eta}{\sqrt{m}} \ltwonorm{\bmu - \bmv} \nonumber \\
	&\leq \frac{1}{2} \Big(CK^2 \frac{\gamma_1 + \gamma_2 + 2\eta}{\sqrt{m}}\Big)^2 + CK^2 \frac{\gamma_1 + \gamma_2 + 2\eta}{\sqrt{m}} \nonumber \\
	&\leq C''K^2 \frac{\gamma_1 + \gamma_2 + 2\eta}{\sqrt{m}},
\end{align*}
where the second inequality follows from the condition of this case and the last inequality holds because of \eqref{sufficientcondition3}.

Combining the above two cases completes the proof.

\end{proof}

\bibliographystyle{IEEEtran}

\bibliography{reference}

%
%
%
%
%
%
%

\end{document}